\documentclass[twocolumn,showpacs,showkeys,superscriptaddress]{revtex4}

\usepackage{amsmath}
\usepackage{bbold}
\usepackage{amsfonts}
\usepackage{amssymb}
\usepackage{pbsi}
\usepackage[T1]{fontenc}
\usepackage{hyperref}
\usepackage{xcolor}
\usepackage{graphicx}
\usepackage{subfig}
\usepackage{float}

\usepackage{color}

\begin{document}

\title{Exact and paraxial Airy propagation of relativistic electron plasma wavepackets}

\author{Maricarmen A. Winkler}
\affiliation{Departamento de F\'isica, Facultad de Ciencias, Universidad de Chile, Casilla 653, Santiago, Chile.}

\author{Camilo V\'asquez-Wilson}
\affiliation{Facultad de Ingenier\'ia y Ciencias,
Universidad Adolfo Ib\'a\~nez, Santiago 7491169, Chile.}

\author{Felipe A. Asenjo}
\email{felipe.asenjo@uai.cl (corresponding author)}
\affiliation{Facultad de Ingenier\'ia y Ciencias,
Universidad Adolfo Ib\'a\~nez, Santiago 7491169, Chile.}

\begin{abstract}
 The different forms of propagation 
of relativistic electron plasma wavepackets in terms of Airy functions are studied. It is shown that  exact solutions can be constructed showing accelerated propagations along coordinates transverse to the thermal speed cone coordinate. Similarly, Airy propagation is a solution for relativistic electron plasma waves in the paraxial approximation. This regime is considered in time-domain, when paraxial approximation is considered for frequency, and in space-domain, when paraxial approximation is considered for wavelength. In both different cases, the wavepackets remains structured in the transverse plane. Using these solutions we are able to define generalized and arbitrary Airy wavepackets for electron plasma waves, depeding on arbitrary spectral functions. Examples of this construction are presented. These electron plasma Airy wavepackets are the most general solutions of this kind.
\end{abstract}

\maketitle

\section{Introduction}

Electron plasma waves are one of the simplest
possible longitudinal modes of wave propagation in plasmas. These modes are ubiquitous at different energy and temperature scales \citep{dawson},
and they are usually solved in terms of linearized plane waves, with a simple dispersion relation. 

However, there are other possible propagation modes of electron plasma wavepackets that have been quite unexplored so far. Taking advantage with similarities with other kind of waves, the purpose of this work is to show that several other relativistic propagation modes for these   exist in terms of Airy functions.  These  modes for electron plasma wavepackets are, in general,  of two different kinds. First, there exist exact three-dimensional spatial wavepacket solutions, that evolve in time and space and exhibit accelerated propagation. Secondly, we also show that there is a second kind of propagating solution, in the paraxial approximation, that can also be obtained in terms of Airy functions. These approximated wavepacket solutions can  be constructed in two different manners, those that can accelerate along a longitudinal direction and others that have curved trajectories in space. It is important to remark that all these solutions are structured in the whole three-dimensional space.
Similar accelerating solutions for both normal and anomalous dispersion of the  paraxial equation derived from Klein-Gordon equation have already been studied in, for example, Refs.~\cite{Eichelkraut1,Eichelkraut2}.

Airy-like solutions are a general form for wave propagation that emerge in different natural phenomena, such as light \citep{neomi,Jiang,abdo,bouch,esat,panag,moya,Baumgartl,Nikolaos,chong, Kaminer}, fluids \citep{water}, sound \citep{chencho,zhao}, gravitational waves \citep{asenjograv}, and heat diffusion \citep{bookolivier,asenjoheat}, among many others.
Therefore, its expected to obtain this kind of propagation also for plasma waves. Some simple Airy waves have been already explored as solutions  \citep{minovich,hehe}. However, and differently to what have been done in previous works, in here we show that 
that a general (exact and paraxial) wavepacket Airy-like solution can be arbitrarily proposed in order to have structured and non-diffracting propagation of these cold electron waves
at  relativistic regimes. With these solutions, even more general structured wavepackets can be constructed in an arbitrary fashion, with the help of arbitrary spectral functions. Therefore, this work proposes complete and new and general solutions for electron plasma waves, not previously considered .

In the following section we  show the derivation of the general equation for relativitsic electron waves. In Secs. III and IV we  thoroughly discuss  the exact and paraxial solutions that allow to construct the Airy wavepackets for electron waves. In Sec. V we show how general and different wavepackets can be constructed using spectral functions.
  Finally, in Sec. VI, we highlight the conclusions of this work.

\section{Relativistic electron plasma waves}

It is very well-known that  a Klein-Gordon equation   describes a Lorentz-invariant cold electron plasma wave. In this section, we will derive such equation, reviewing the important assumptions that lead to it.
 By using a general covariant formalism for a 
 relativistic  multifluid plasma, we start by deriving the wave equations for electron plasma waves and for electromagnetic plasma waves in the cold limit, when the thermal energy is much less than the rest-mass energy of the plasma constituents. The momentum equation for $j$ species is
\begin{equation}
    \partial_\nu \left(m_j n_j U^\mu_j U^\nu_j+p_j \eta^{\mu\nu}\right)=q_j n_j F^{\mu\nu}U_{\nu}^j\, ,
    \label{momentumcovariant}
\end{equation}
where $F^{\mu\nu}$ is the electromagnetic tensor,  with charge $q_j$, mass $m_j$, pressure $p_j$, density in the rest-frame $n_j$, and four-velocity $U_j^\mu=(\gamma_j,\gamma_j {\bf v}_j)$, where ${\bf v}_j$ is the velocity, and $\gamma_j=(1-{\bf v}_j\cdot {\bf v}_j)^{-1/2}$ is its corresponding Lorentz factor. The electromagnetic tensor $F^{\mu\nu}=\partial^\mu A^\nu-\partial^\nu A^\mu$ is written in terms of the electromagnetic potential $A^\mu$.
The above equation is considered in a flat spacetime with metric $\eta^{\mu\nu}=\mbox{diag}(-1,1,1,1)$ \citep{misner}. 
The system is finally complemented by the Maxwell equations
\begin{equation}
    \partial_\nu F^{\mu\nu}=\sum_j q_j n_j U^\mu_j\, ,
    \label{Maxweelcovariant}
\end{equation}
which implies the conservation of charge for each plasma component
\begin{equation}
    \partial_\mu \left(  q_j n_j U^\mu_j\right)=0\, .
    \label{conservationcovariant}
\end{equation}

We can use the above general theory to study relativistic electron waves. 
In this case,
let us assume a fixed ion fluid background (considered at rest), with a mobile electron fluid with charge $-e$ and mass $m$. 
In order to study the dynamic of this mode, we will perform a perturbation scheme. Thus, the electron fluid
has a rest-frame density $n=n_0+{\tilde n}$, where $n_0$ is the constant background density, and ${\tilde n}\ll n_0$ is the perturbed density. Let us consider that the electron plasma with a perturbed velocity  ${\bf v}= {\tilde{\bf v}}$.  Therefore, the electron fluid relativistic Lorentz factor is
${\gamma}\approx 1$.
On the other hand, the electron plasma is immersed in a  perturbed  electrostatic field ${\tilde {\bf E}}$.  In such case, the Gauss' Law \eqref{Maxweelcovariant} reduces to 
\begin{equation}
    \nabla\cdot {\tilde {\bf E}}=-e {\tilde n}\, .
    \label{equation1ini0}
\end{equation}
Also, the electron fluid fulfill the conservation law \eqref{conservationcovariant}
\begin{equation}
    \frac{\partial {\tilde n}}{\partial t}+n_0\nabla\cdot{\tilde {\bf v}}=0\, .
    \label{equation2ini0}
\end{equation}
Lastly, 
the electron fluid momentum equation becomes \eqref{momentumcovariant}
\begin{equation}
    m \frac{\partial{\tilde {\bf v}}}{\partial t}=-e\, {\tilde {\bf E}}-\frac{m S^2}{n_0}\nabla{\tilde n}\, , 
    \label{equation3ini0}
\end{equation}
where  $S^2=(1/m)\partial {\tilde p}/\partial {\tilde n}$ is the electron thermal speed, with pertubative pressure ${\tilde p}$. The thermal speed depends on temperature, $S=S(T)$, such that $k_B T\ll m c^2$. 
Using Eqs.~\eqref{equation1ini0}-\eqref{equation3ini0} we finally are able to obtain the equation for relativistic dynamics of electron plasma density 
\begin{equation}
    \left(\frac{\partial^2}{\partial t^2} - {S^2} \nabla^2+{\omega_p^2} \right){\tilde n}=0\, ,
    \label{densityequation}
\end{equation}
where  $\omega_p=\sqrt{e^2 n_0/m}$ is the electron plasma frequency.

  Whenever some of the previous assumptions are relaxed, such as  considering background velocities or relativistic temperatures, then a more general and complete form of Eq.~\eqref{densityequation} can be found for arbitrary background electron velocities. A detailed study of this is in Ref~\citep{fasenjovv}.

  Eq.~\eqref{densityequation} is fully relativistic in this perturbative scheme.
It is a common practice to solve it  in terms of plane waves, obtaining thus a simple dispersion relation for different modes. However, by similarities with wave equations in other fields, we know that
Airy wavepackets are also propagating solutions of Eq.~\eqref{densityequation} in an exact manner and in the paraxial wave approximation. We show here that these both Airy solutions are different between them, as they have different propagation characteristics. Therefore, we are able to show how these solutions can be used to construct general electron plasma wavepackets with arbitrary transversal forms.

\section{Exact propagation of Airy electron plasma  wavepacket}
\label{exactsoluAiry}

Let us start by showing that Eq.~\eqref{densityequation} can be solved exactly in terms of an Airy-like propagating wave. Usually,  Eq.~\eqref{densityequation}  is solved in the paraxial limit to show its Airy-like properties. This will be done in the following sections.  However, as Eq.~\eqref{densityequation} is a relativistic equation (obtained from a covariant formalism) and therefore it can be solved by considering time and spatial coordinates in the same footing. Thus, the exact wavepacket  can be solved in terms of the characteristic thermal speed cone coordinates $x\pm S t$. In this way, this wavepacket accelerates in the transversal direction to such coordinates. 
 In order to show this, let us consider the following form of the perturbed density
\begin{equation}
    {\tilde n}(t, {\bar{\bf x}})=\psi_0\,  \zeta(\eta, { y},{ z})\exp\left(i\alpha\, \xi-i\frac{\omega_p^2}{4\alpha S^2}\eta \right)\, ,
    \label{exactnairprogrpab}
\end{equation}
where $\eta=x+S t$,  $\xi=x-S t$ \citep{imb},  $\alpha$ is an arbitrary constant with inverse length unit  dimensions, and ${\tilde n}_0$ is an arbitrary constant.
Using this in 
Eq.~\eqref{densityequation}, we obtain that the function $\zeta$ follows a Schr\"odinger-like equation
\begin{equation}
    i\frac{\partial \zeta}{\partial \eta}=-\frac{1}{4\alpha}\left(\frac{\partial^2}{\partial { y}^2}+\frac{\partial^2}{\partial { z}^2} \right)\zeta\, ,
\end{equation}
which can be solved by Airy functions ${\mbox{Ai}}$. Following Lekner's results
\citep{lekner}, a normalizable wavepacket solution to the above equation can be written as $\zeta(\eta,y,z)=\zeta_y(\eta,y)\zeta_z(\eta,z)$, where
\begin{eqnarray}
\zeta_{y,z}&=&{\mbox{Ai}}\left[(8 a_{y,z} \alpha^2)^{1/3}\left(w_{y,z}-u_{y,z} \eta+i v_{y,z} \eta- a_{y,z} \frac{\eta^2}{2}\right)\right]\nonumber\\
&&\times\exp\left[2i \alpha a_{y,z} \eta\left( w_{y,z}-u_{y,z} \eta-a_{y,z} \frac{\eta^2}{3} \right)  \right]\nonumber\\
&& \times\exp\left[2\alpha v_{y,z}\left(w_{y,z}-u_{y,z} \eta+i v_{y,z} \frac{\eta}{2}-a_{y,z} \eta^2 \right) \right]\nonumber\\
&&\times\exp\left[2i\alpha u_{y,z}\left(w_{y,z}-u_{y,z} \frac{\eta}{2} \right) \right]\, ,
\end{eqnarray}
with $w_y\equiv y$, $w_z\equiv z$,  $u_{y,z}$ 
are arbitrary effective Galilean boost speeds in respective $y$ and $z$-directions, 
and $v_{y,z}>0$ are arbitrary factors that allow the solution to be integrable. Both, $v_y$ and $v_z$
 are parameters that allow to determine how fast the solution decays at infinity \citep{lekner}.
 Besides, $a_{y,z}$ are arbitrary accelerations in respective $y$ and $z$-directions. 

In this way, the respective relativistic electron plasma wavepacket can propagate in an exact manner in terms of Airy functions, showing that it has different high-intensity lobes, presenting independent accelerations in $y$ and $z$
directions with respect to the $\eta$ direction. Of course, another straightforward solution exists that has acceleration   in the transverse plane with respect to the $\xi$ direction. 

\begin{figure*}
\subfloat[]{\includegraphics[width = 1.8in]{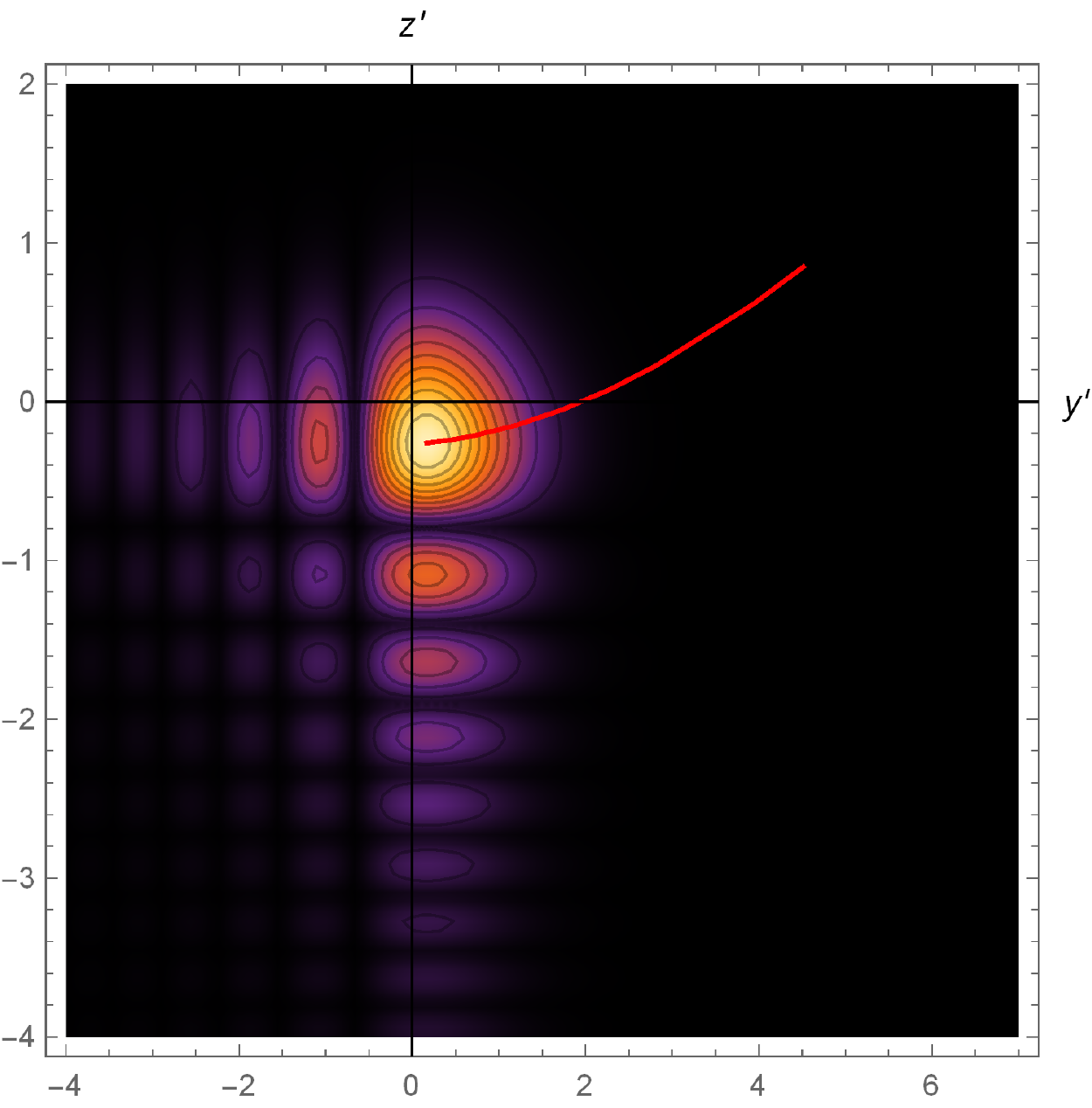}} 
 \subfloat[]{\includegraphics[width = 1.8in]{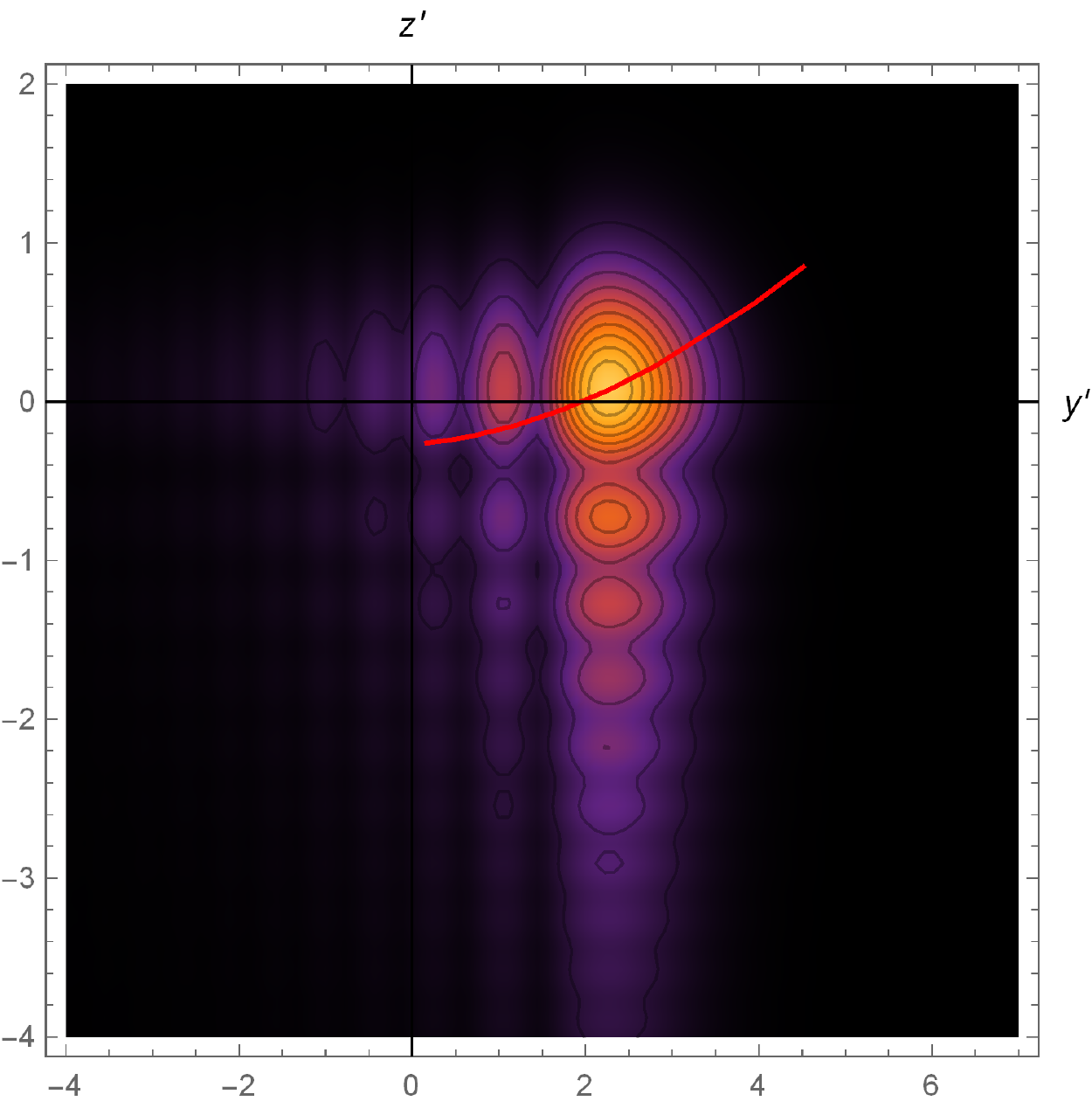}}
\subfloat[] {\includegraphics[width = 1.8in]{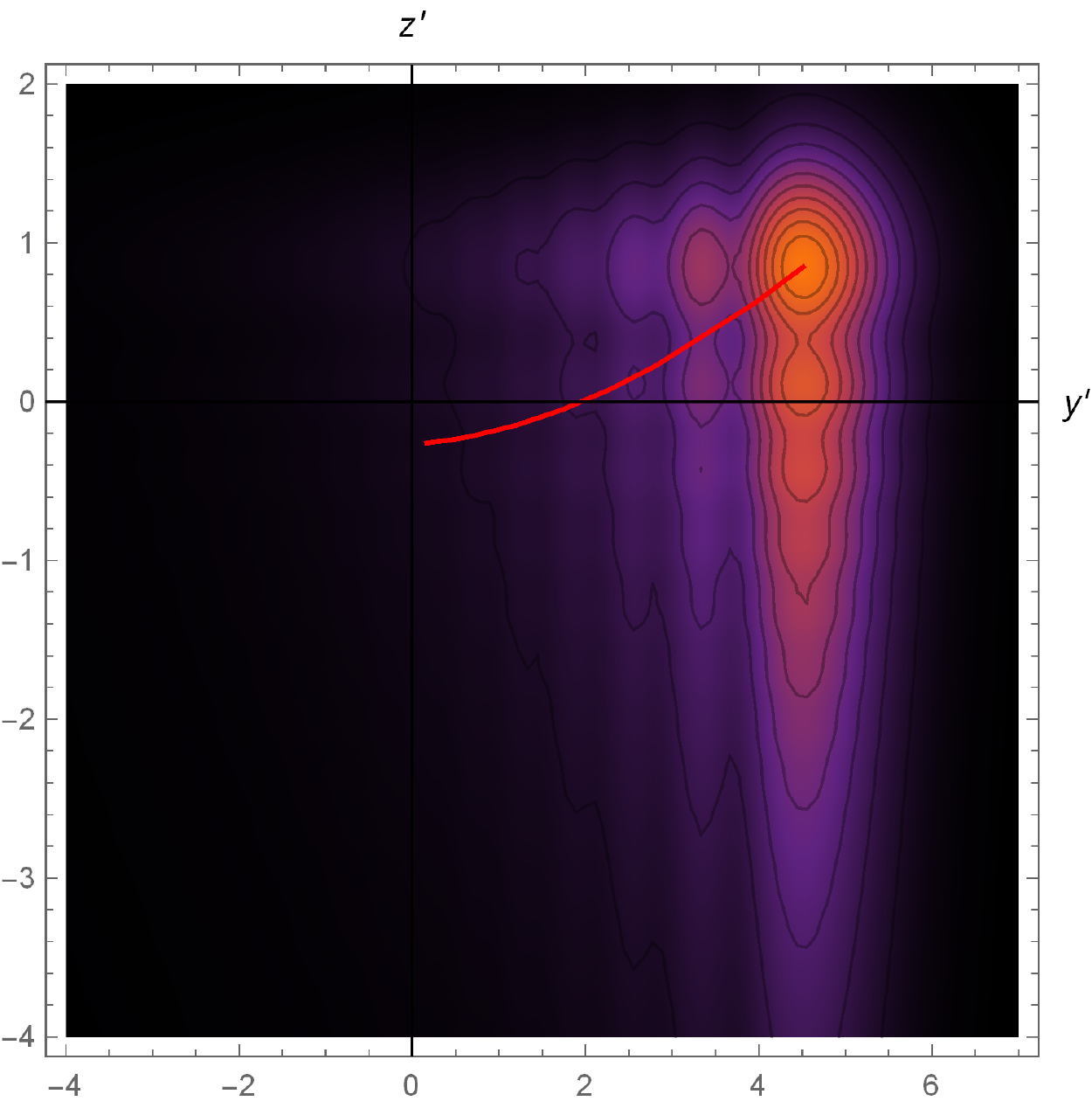}}\\
\subfloat[]{\includegraphics[width = 1.8in]{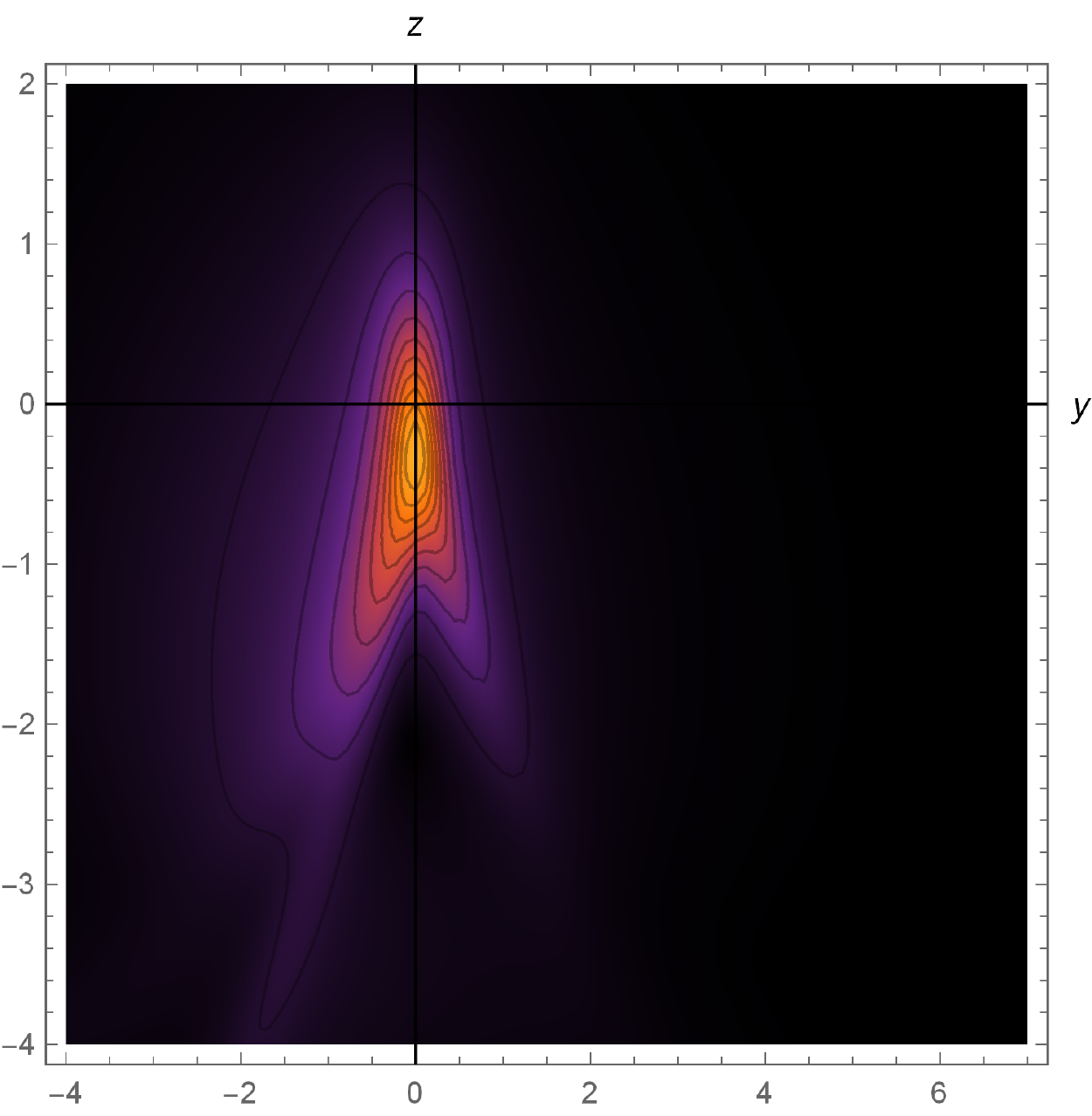}} 
 \subfloat[]{\includegraphics[width = 1.8in]{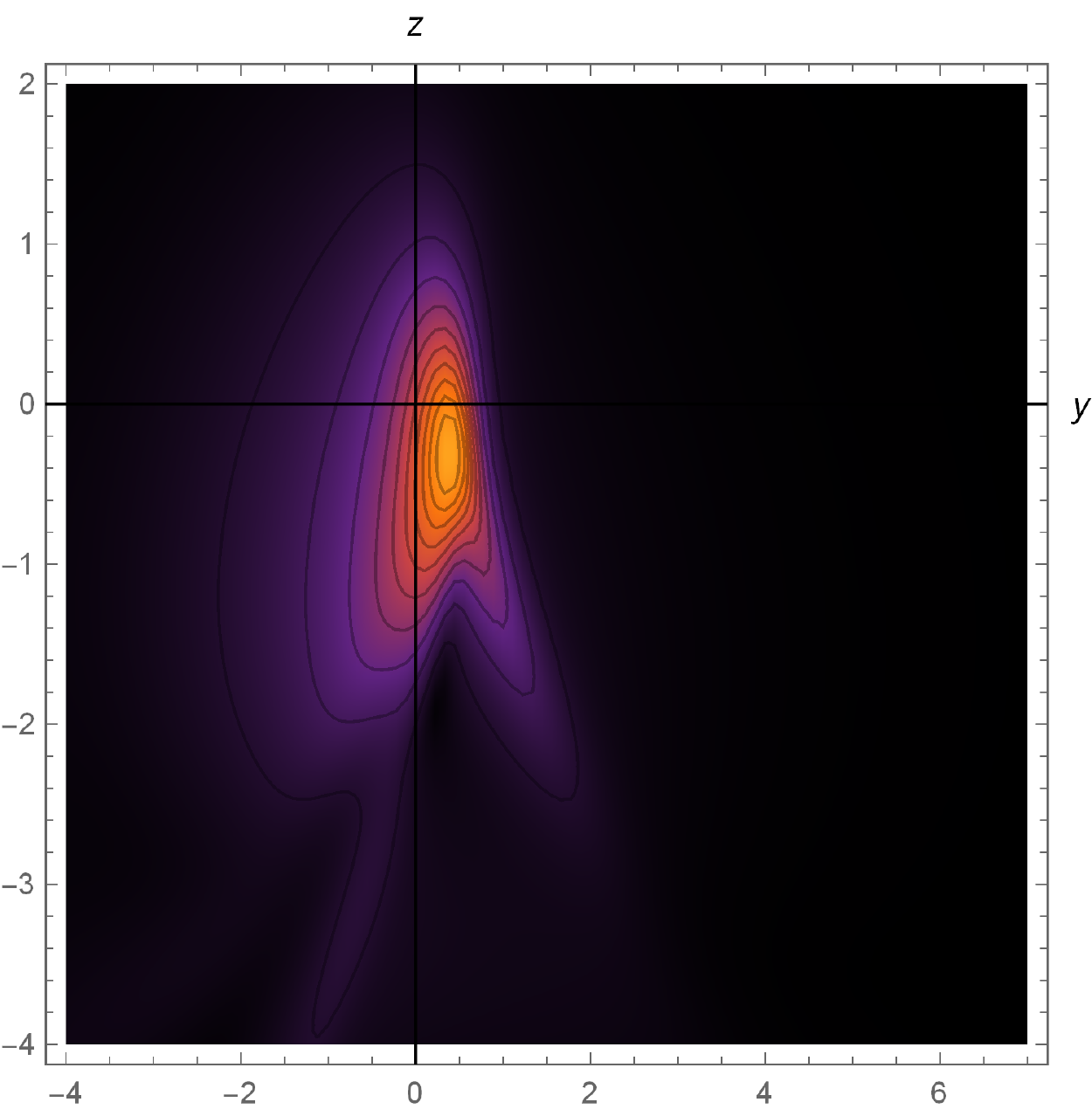}}
\subfloat[] {\includegraphics[width = 1.8in]{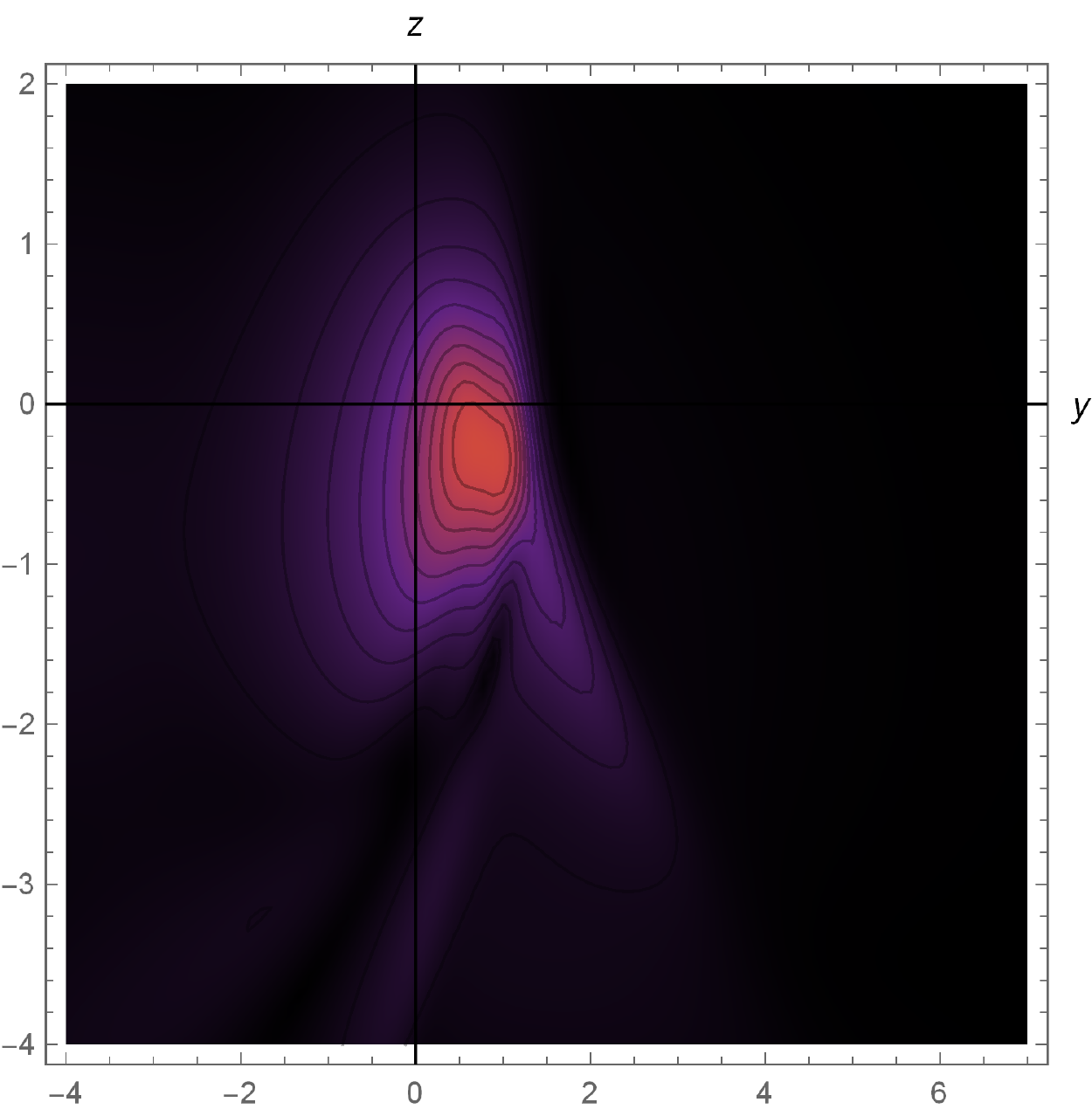}}\\
\caption{Different evolution density plots for magnitude of densities $|{\tilde n}|$ solutions. Countours plots are also shown for different solutions. Brighter colors stand for higher density values. Figs. (a), (b) and (c) are  solution \eqref{exactnairprogrpab}, for $\alpha\eta=0.1$, $\alpha\eta=0.5$, and $\alpha\eta=1$, respectively. In red line, we show the total parabolic trajectory of the maximum maximorum of the density for all $\eta$.  On the other hand, Figs. (d), (e) and (f) are for  the magnitude of density $|{\tilde n}_G|$ of the generalized wavepacket solution of Sec.~\ref{generlaAiryformwavepa}, for $g(\alpha)=\exp(-\alpha^2)$, and
 for $\eta=0.05$, $\eta=0.2$, and $\eta=0.4$, respectively.}
\label{figura1}
\end{figure*}

As an example, the above complete exact solution is shown in Figs.~\ref{figura1}(a), (b) and (c) for $|{\tilde n}|$. We have used normalized coordinates $y'=\alpha y$ and $z'=\alpha z$.   In order to exemplify the general behavior of the solution, we have used arbitrary
parameter values $v_y=v_z=0.3$,
$u_y=5$, $u_z=0$,   and normalized acceleration $a_y/\alpha=1$ and $a_z/\alpha=3$. The dynamics is considered for three different normalized  speed coordinates, $\alpha\eta=0.1$ for Fig.~\ref{figura1}(a), $\alpha\eta=0.5$ for Fig.~\ref{figura1}(b), and $\alpha\eta=1.5$ for Fig.~\ref{figura1}(c). We can notice how the density profile follow curved trajectories in $y-z$ space as $\eta$ increases. This is shown in a  red line, that display the total numerical parabolic trajectory calculated for the maximum maximorum of the density.  For this solution,  different values of $a_{y,z}$ allows to have different curved trayectories in the $y-z$ plane, while the wave propagates along the $\eta$ direction.

The above presented exact solution differs from previous results for Airy electron waves \citep{hehe}, where only a paraxial approximation in space was considered for non-relativistic plasmas.   In the current case, this solution is a consequence of
the balance of space and time, because of the
 relativistic four-dimensional wave equation \eqref{densityequation}.

Differently, in the following section we  show that waves in the paraxial approximation in time or in space also can be solved in terms of Airy function for relativistic electron plasmas.

\section{Paraxial propagation of  Airy electron plasma wavepackets}

Differently to the previous exact solution, we can also show that in the paraxial approximation, structured, localized and non-diffracting Airy-like propagation solutions exist. In fact, different behaviors are found depending on the kind of paraxial approximation used, say in time-domain or in space-domain. In the following, we discuss these two types of solutions, showing their different expected behaviors.

\subsection{Accelerating  wavepackets}
\label{paraxialsubsec1}

In the paraxial approximation one can also   obtain  solutions that propagate in accelerated fashion in time as they move in space. In this case, the paraxial approximation should be performed in the time-domain of the solution.

In order to solve  Eq.~\eqref{densityequation} under this assumption, let us consider a solution for electron density ${\tilde n}$ with slowly-varying dependence $\rho$, and a rapid-varying phase, with frequency $\omega$, in the form
\begin{equation}
    {\tilde n}(t,{\bf x})= {\tilde n}_0\, \rho (t,{\bf x})\, \exp(i\omega t)\, ,
    \label{ntimedomain}
\end{equation}
where ${\tilde n}_0$ is a constant.
The paraxial approximation in time-domain can be taken when $\omega\gg \partial_t^2\rho/\partial_t \rho$ is considered. This occurs when the time variation scale of $\rho$  is much larger than the oscillation time scale of the wave $1/\omega$.

In this way, using \eqref{ntimedomain},  Eq.~\eqref{densityequation}
is written for $\rho$ as 
\begin{equation}
    \left(2i\omega\frac{\partial}{\partial t}  - S^2\,\nabla^2+{\omega_p^2} -\omega^2\right){\rho}=0\, .
    \label{densityequationparaxial1}
\end{equation}
Now, let us consider an electron plasma wavepacket propagating in a longitudinal direction, say $z$, where in the transverse direction it remains with a structure described by Bessel functions $J_n$. In that case, 
\begin{eqnarray}
    \rho (t,{\bf x})&=&\rho_z(t,z) J_n(\alpha r)\nonumber\\
    &&
   \exp\left(i n \phi-i\frac{\omega^2-S^2\alpha^2 -\omega_p^2}{2\omega}t\right)\, ,
   \label{solrhozJ}
\end{eqnarray}
where $r=\sqrt{x^2+y^2}$,  $\phi=\arctan(y/x)$, and $\alpha$ is a constant. Using this ansatz in Eq.~\eqref{densityequationparaxial1}, we can obtain  an expression for the dynamics of $\rho_z$, which becomes simply
\begin{equation}
    i\frac{\partial \rho_z}{\partial t}=\left(\frac{S^2}{2  \omega}\right)\frac{\partial^2 \rho_z}{\partial z^2}\, .
\end{equation}
This last equation can again be solved in terms of Airy functions \citep{berry,lekner} to have the normalizable solution
\begin{eqnarray}
\rho_{z}(t,z)&=&{\mbox{Ai}}\left[\left(2 a \frac{ \omega^2}{S^4}\right)^{1/3}\left(z-u t-i v t- a \frac{t^2}{2}\right)\right]\times\nonumber\\
&&\exp\left(-i \frac{\omega}{S^2}a t\left( z-u t-a  \frac{t^2}{3} \right)  \right)\times\nonumber\\
&& \exp\left( \frac{ \omega}{S^2} v\left(z-u t-i v  \frac{t}{2}-a t^2 \right) \right)\times\nonumber\\
&&\exp\left(-i \frac{\omega}{S^2} u\left(z-u  \frac{t}{2} \right) \right)\, ,
\end{eqnarray}
where $u$ is a Galilean boost speed along $z$-direction \citep{lekner}, $v$ is an arbitrary factor that enables the normalization, and $a$ is the arbitrary acceleration of the solution  in the $z$--$t$ plane.

\begin{figure*}
\subfloat[] {\includegraphics[width = 1.8in]{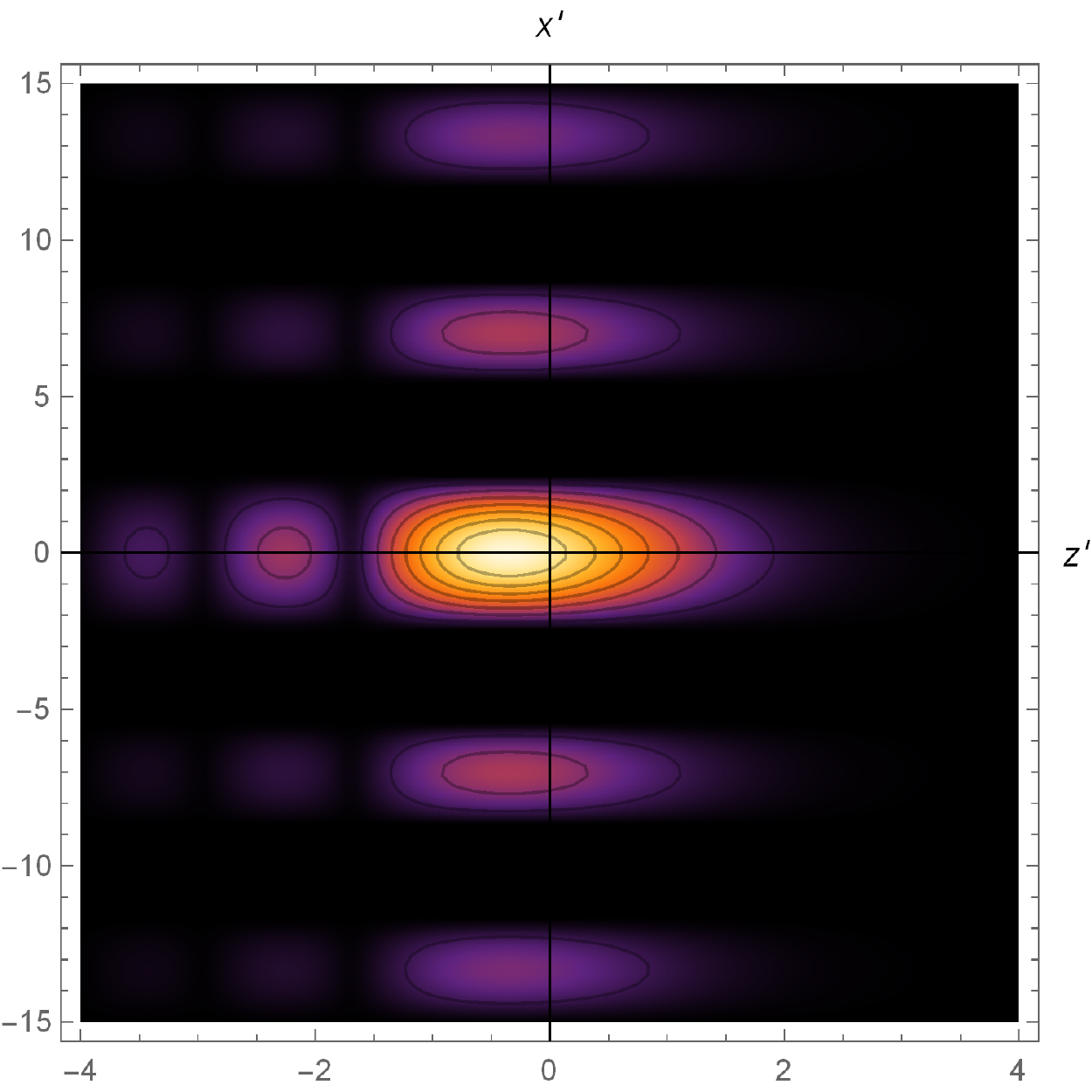}} 
\subfloat[] {\includegraphics[width = 1.8in]{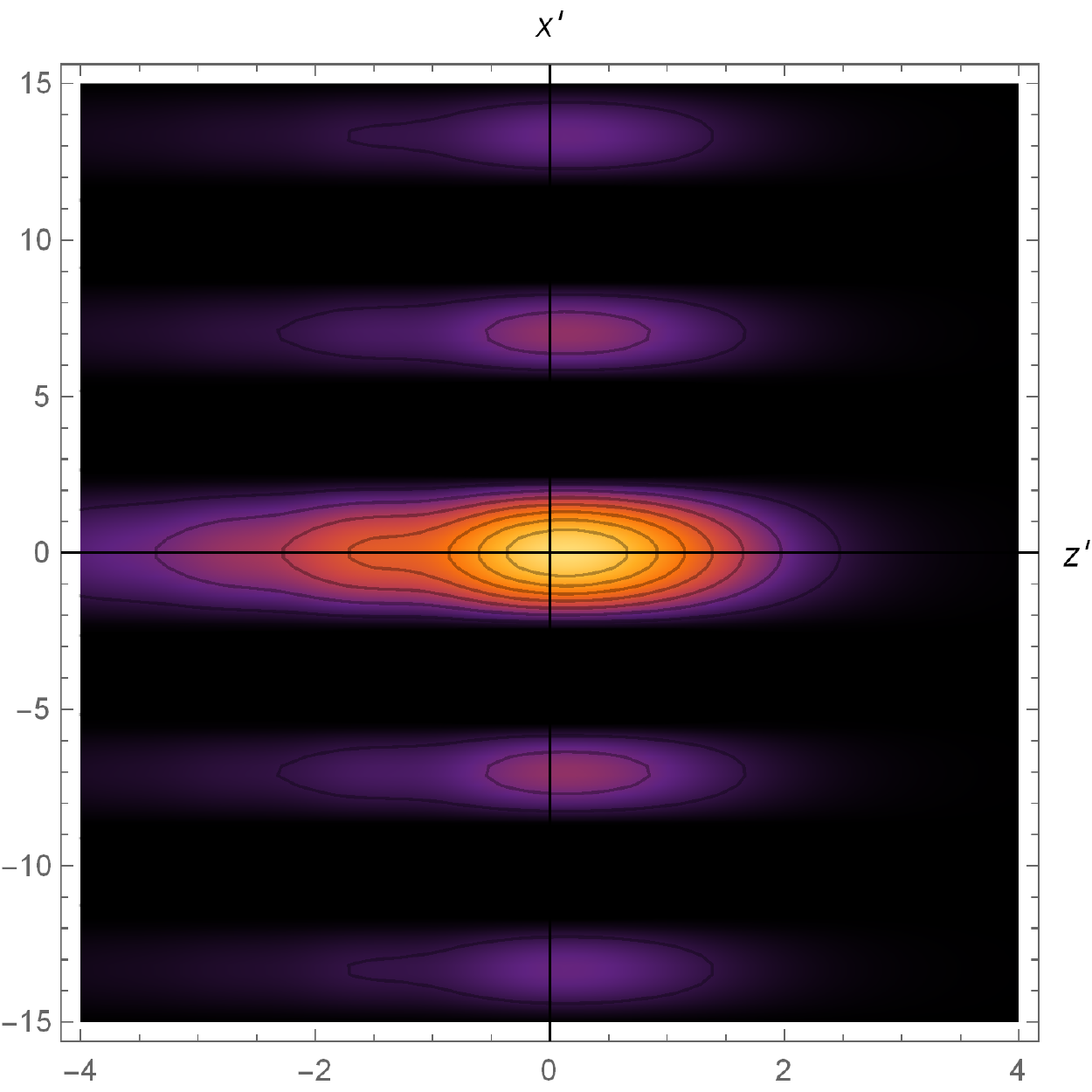}} 
\subfloat[] {\includegraphics[width = 1.8in]{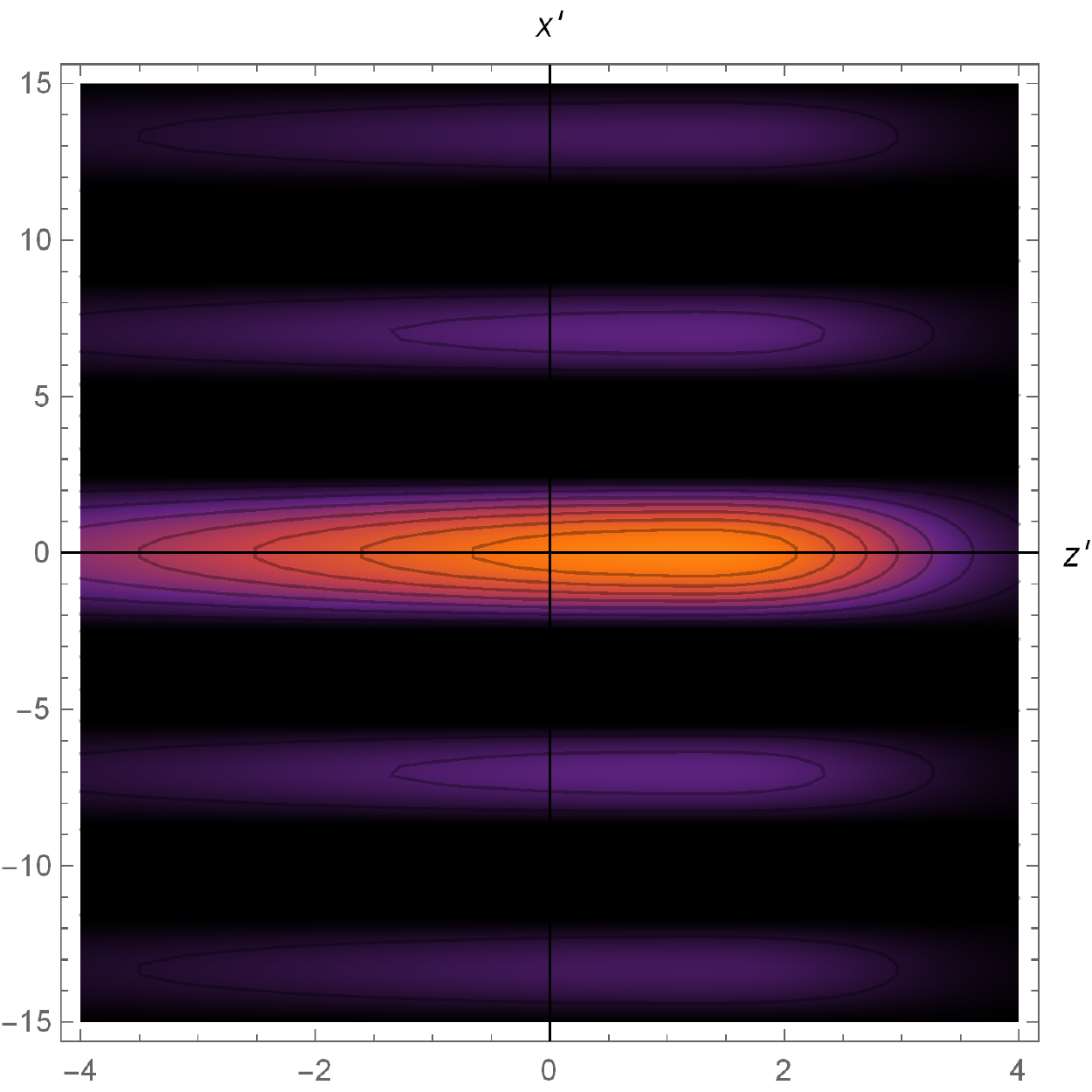}}\\ 
\subfloat[] {\includegraphics[width = 1.8in]{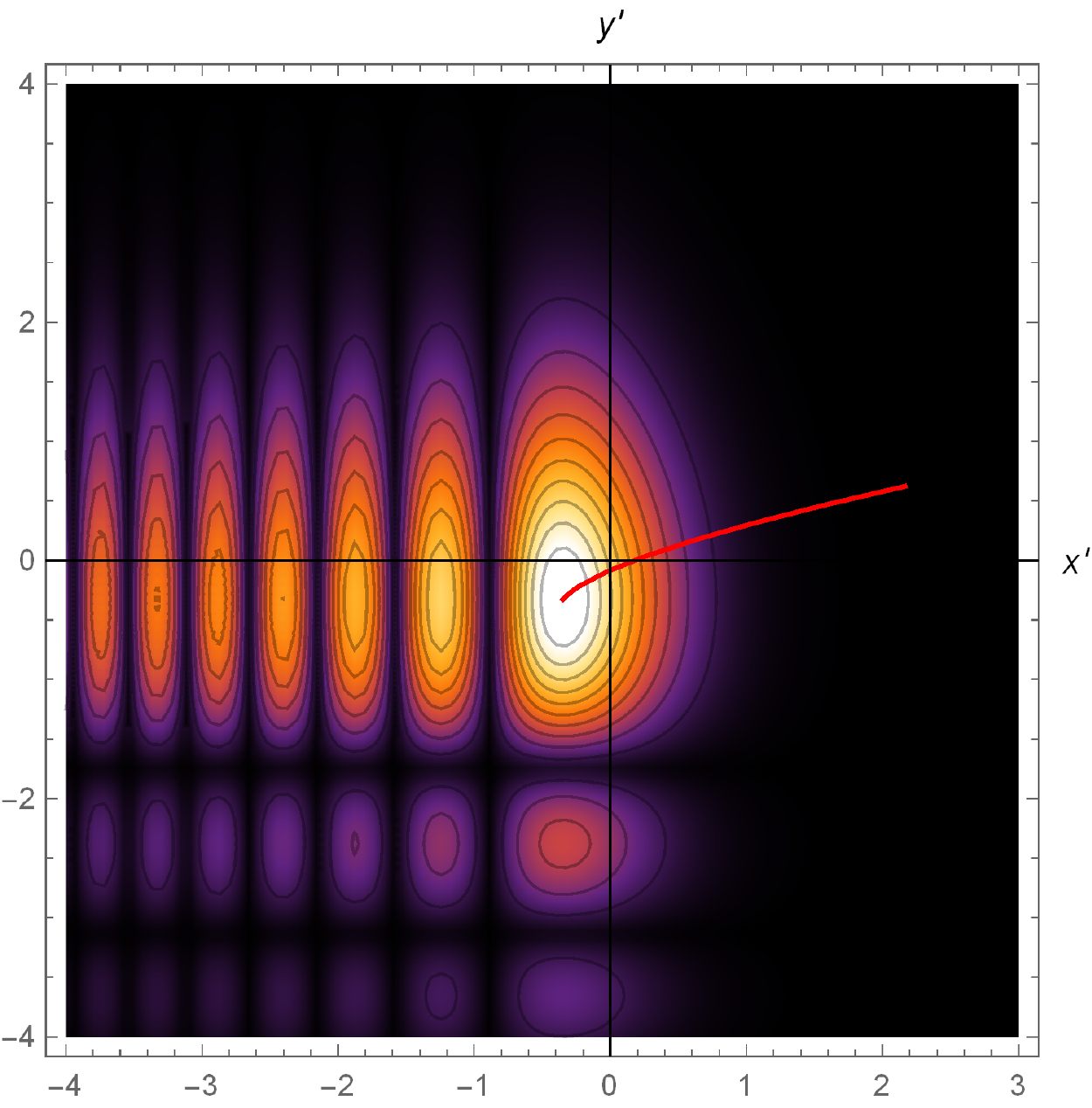}} 
\subfloat[] {\includegraphics[width = 1.8in]{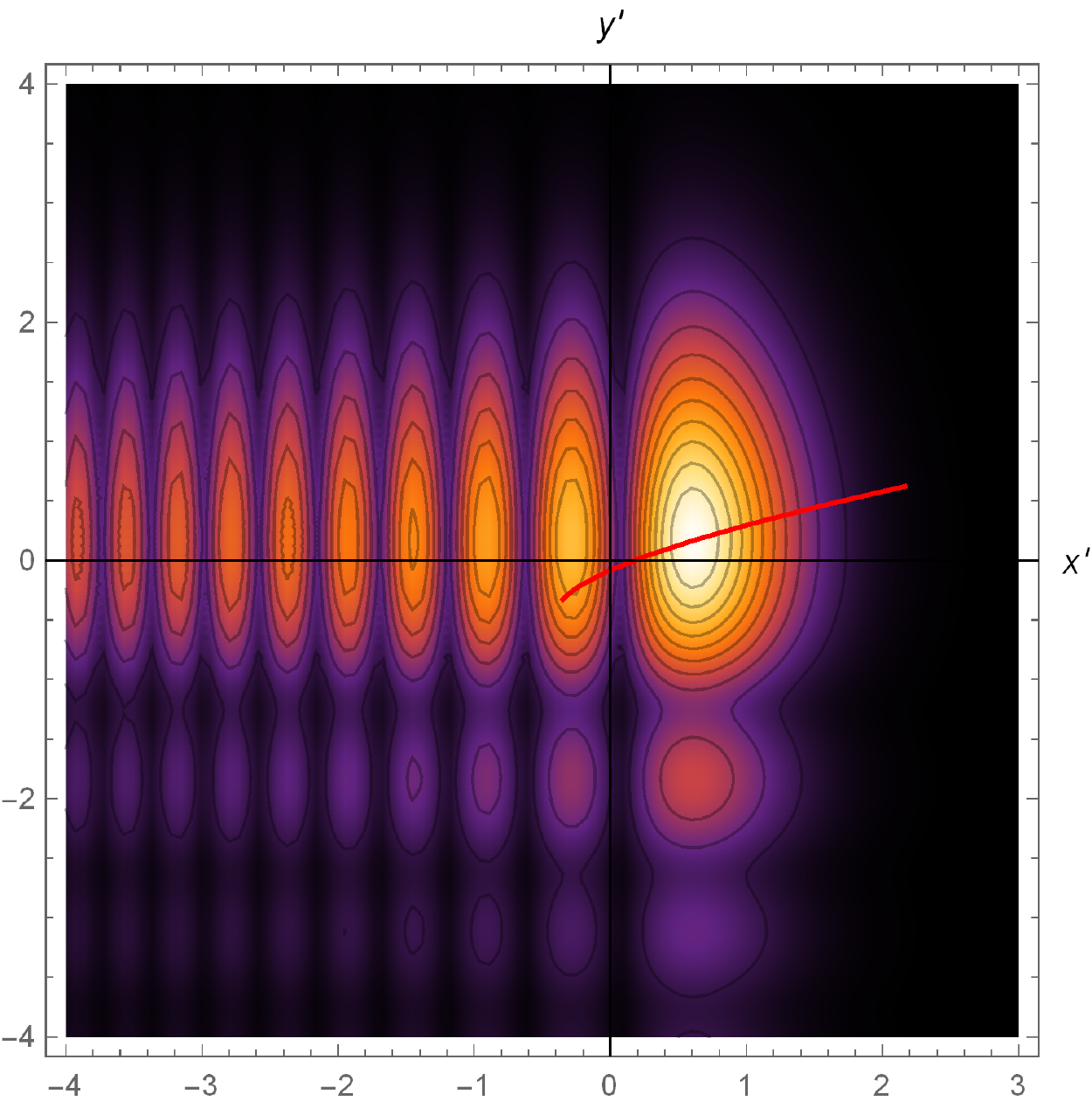}} 
\subfloat[] {\includegraphics[width = 1.8in]{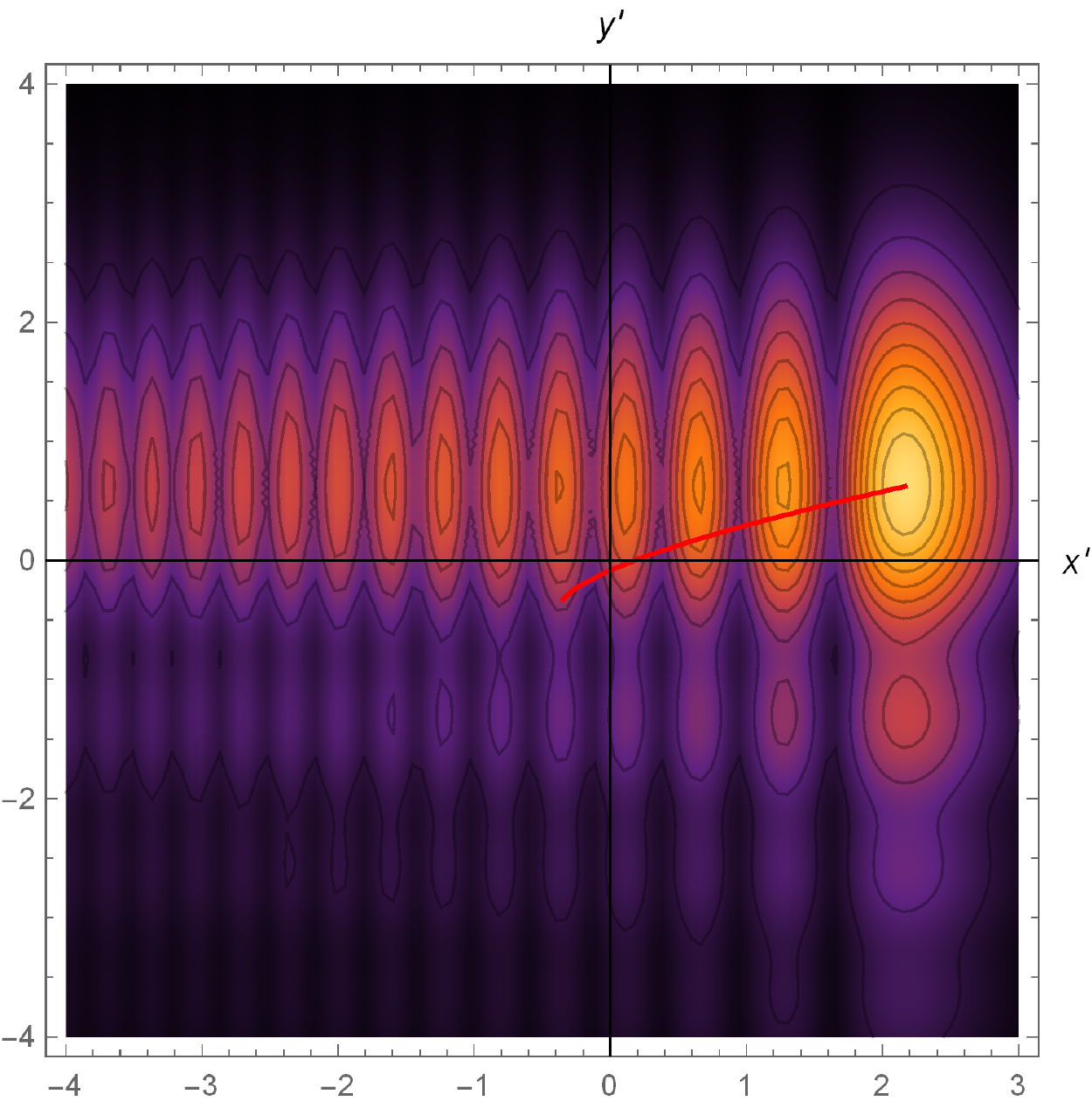}} 
\caption{Different evolution density plots for magnitude of densities $|{\tilde n}|$ solutions. Countours plots are also shown for different solutions. Brighter colors stand for higher density values. 
Figs.~\ref{figura1bb}(a), (b) and (c)
are  density solution \eqref{ntimedomain}, for 
$\omega_p t=0.1$, $\omega_p t=1$, and $\omega_p t=2$, respectively.   Figs.~\ref{figura1bb}(d), (e) and (f) are density solution \eqref{ntimedomainspace}, for $kz=0.1$, $kz=0.5$, and $kz=0.8$, respectively. In red line, we show the parabolic trajectory of the maximum maximorum of the density for all $z$.}
\label{figura1bb}
\end{figure*}

As a proper example for this behavior, the magnitude of the density solution \eqref{ntimedomain} is plotted in Figs.~\ref{figura1bb}(a), (b) and (c).   For this,  we have  normalized the velocity to the inertial and thermal  responses of the plasma wave, $\sqrt{\omega/\omega_p}\, v/S$. Similar normalization is choosen for  acceleration, $\sqrt{\omega/\omega_p}\,  a/(S\omega_p)$. This last parameter,  besides of measuring the acceleration of wavepacket along the longitudinal direction,
determines
 the strength of the maximum value of the Airy function, and thus, the maximum value of $\rho_z$.
Thus, the plots are constructed for $u=0$, normalized factor $\sqrt{\omega/\omega_p}\, v/S=0.6$,
and normalized acceleration $\sqrt{\omega/\omega_p}\,  a/(S\omega_p)=1.3$. 
The evolution is considered for normalized times 
$\omega_p  t=0.1$ in Fig.~\ref{figura1bb}(a), $\omega_p  t=1$ in Fig.~\ref{figura1bb}(b), and $\omega_p  t=2$ in Fig.~\ref{figura1bb}(c). All plots are in terms of normalized transversal coordinates $x'=\alpha x$, $y'=\alpha y$, and normalized longitudinal distance $z'=\sqrt{\omega\omega_p}\, z/S_e$. 
The transversal part of the solution follow a Bessel form, whereas it propagates in $z$-direction.

\subsection{Curved trajectory  wavepackets}
\label{paraxialsubsec2}

We can also perform a different paraxial approximation, now in space-domain. This solution consists in wavepackets following curved parabolic trajectories in space, as we show in the following. 

Let us consider that the solution of Eq.~\eqref{densityequation} can be put in the form
\begin{equation}
    {\tilde n}(t,{\bf x})=  {\tilde n}_0\, \rho(x,y,z)\, \exp(i\omega t+ik z)\, .
    \label{ntimedomainspace}
\end{equation}
where $ {\tilde n}_0$ is anew an arbitrary constant, $\omega$ is a constant frequency, and the  paraxial
approximation in space-domain is obtained when $k\gg \partial_z^2\rho/\partial_z\rho$. This paraxial limit occurs when the length variation scale of $\rho$ is much larger than the constant wavelength scale of the wave $1/k$.

In this case, using Eq.~\eqref{densityequation},
we arrive to the equation
\begin{equation}
    i\frac{\partial\rho}{\partial z}=-\frac{1}{2k}\left(\frac{\partial^2}{\partial x^2}+\frac{\partial^2}{\partial y^2} \right)\rho+\frac{\beta^2}{2k}\rho\, ,
    \label{equarhospacedomain}
\end{equation}
where we have defined the constant $\beta^2\equiv k^2+(\omega_p^2-\omega^2)/S^2$. Notice that constant $\beta$ is arbitrary and it defines a dispersion relation for the frequency. In principle, $\beta$ can be chosen to  be $0$ (as it is done in Ref.~\cite{hehe} for the non-relativistic cold electron plasma case). However, as we will show in the following Section, a more general wavepacket solution can be constructed from its arbitrary value.

Equation \eqref{equarhospacedomain} has wavepacket solutions that show arbitrary parabolic trajectories in the $x$--$z$ and $y$--$z$ planes \citep{lekner,berry}.  Its solution is given in terms of $\rho(x,y,z,\beta)=\rho_x(x,z)\rho_y(y,z)\exp(-i\beta^2 z/2k)$, with
\begin{eqnarray}
\rho_{x,y}&=&{\mbox{Ai}}\left[(2 a_{x,y} k^2)^{1/3}\left(w_{x,y}-u_{x,y} z+i v_{x,y} z- a_{x,y} \frac{z^2}{2}\right)\right]\nonumber\\
&&\times\exp\left(i k a_{x,y} z\left( w_{x,y}-u_{x,y} z-a_{x,y} \frac{z^2}{3} \right)  \right)\nonumber\\
&&\times \exp\left(k v_{x,y}\left(w_{x,y}-u_{x,y} z+i v_{x,y} \frac{z}{2}-a_{x,y} z^2 \right) \right)\nonumber\\
&&\times\exp\left(i k u_{x,y}\left(w_{x,y}-u_{x,y} \frac{z}{2} \right) \right)\, ,
\end{eqnarray}
where $w_x\equiv x$, $w_y\equiv y$,   and $u_{x,y}$ 
are arbitrary effective Galilean boost speeds in $x$ or $y$ directions. Once again, $v_{x,y}>0$ produces  an integrable solution. 
Similar to previous cases, $a_{x,y}$ are the accelerations of the wavepacket along the $x$--$z$ and $y$--$z$ planes, respectively. 

  This solution is a relativistic generalization of the  results presented in Ref.~\cite{hehe}. The  differences are that each transverse direction ($x$ and $y$) can accelerate in an independent fashion. Also, the relativistic solution allows to have a finite energy wavepacket (by including velocities $v_{x,y}$), being localized in space, which in principle is not forced to follow the dispersion relation of a classical electron plasma wave, $\beta\neq 0$. This is crucial when generalized wavepacket are defined, as it is shown in Sec.~\ref{gencurvrajeplaswavep}.

The evolution of the magnitude of the density \eqref{ntimedomainspace} is plotted in Figs.~\ref{figura1bb}(d), (e) and (f), in 
terms of normalized coordinates $x'=kx$, $y'=ky$, and $z'=kz$. 
For this case, we have used the normalized accelerations $a_x/k=8$, $a_y/k=1$, and $u_x=0$, $u_y=1$, $v_x=0.1$ and $v_y=1/2$. This parameters allow us to show the curvature trajectory in the $x-y$ plane, for different $z$ values.
The Fig.~\ref{figura1bb}(d) is for $z'=0.1$, 
Fig.~\ref{figura1bb}(e) is for $z'=0.5$, 
and Fig.~\ref{figura1bb}(f) is for $z'=0.8$. The figures show that the density of the electron plasma wave propagates following curved  arbitrarily independent trajectories in the $x-z$ and $y-z$ planes. 
The followed parabolic trajectory of the maximum maximorum of the magnitude of the density
is also depicted  as a red line in the plots.

  For this solution, importantly, the acceleration parameters $a_{x,y}$ determine the curvature of the trajectory of the wave along the $z$ direction of proapagtion, as well as thr strenght of the maximum of the Airy solution.

\section{Generalized electron plasma wavepackets}

The above analysis  of Secs.~\ref{exactsoluAiry}, \ref{paraxialsubsec1} and \ref{paraxialsubsec2} shows exact and paraxial solutions
that depend on arbitrary constants, $\alpha$ and $\beta$ respectively. Using these constant, generalized wave packets can be constructed by the introduction of spectral functions. These functions works of weights for an averaged general wavepacket.

\subsection{General form for exact Airy electron plasma wavepacket} 
\label{generlaAiryformwavepa}

The above solution \eqref{exactnairprogrpab} can be used to construct the most general solution for this kind of wavepackets. 
This can be done by constructing the wavepacket as an average of previous solution through a function that depends only on the introduced arbitrary parameter $\alpha$. Thus, a general accelerating relativistic plasma density wavepacket ${\tilde n}_G$
can be written as \citep{iwo}
\begin{eqnarray}
    {\tilde n}_G(t,{\bf x})&=& \int d\alpha\,  g(\alpha) {\tilde n}(t,{\bf x},\alpha)\, ,
\label{ildengeneralejem1}
\end{eqnarray}
where ${\tilde n}(t,{\bf x},\alpha)$ is solution \eqref{exactnairprogrpab},  and $g(\alpha)$ is an arbitrary spectral function, that can be chosen at will to define different wavepacket structures. Clearly, wavepacket \eqref{ildengeneralejem1} is solution of Eq.~\eqref{densityequation}.

To exemplify the construction of this general wavepacket, let us consider the  case 
when $g(\alpha)=\exp(-\alpha^2)$. The  magnitude  of density $|{\tilde n}_G|$ for this case  is presented in Figs.~\ref{figura1}(d), (e) and (f). The wavepacket is obatined by numerical integration for positive values of $\alpha$,  and for
    arbitrary
parameter values $v_y=v_z=0.3$,
$u_y=5$, $u_z=0$, $a_y=1$ and $a_z=3$. The three figures are considered in the plane $\xi=1$, while $\eta=0.05$ for Fig.~\ref{figura1}(d),  $\eta=0.2$ for Fig.~\ref{figura1}(e), and  $\eta=0.4$ for Fig.~\ref{figura1}(f). 

Notice how this wavepacket gets more compacted comapred with solution \eqref{exactnairprogrpab} [shown in  Figs.~\ref{figura1}(a), (b) and (c)], implying that the maximum intensity of this solution is now more localized. The curved trajectory of the propagation, however, remains conserved.
Other possible choices of $g(\alpha)$ can produce different behavior for propagation.

\subsection{General form for accelerating plasma wavepacket} 

Similarly to the previous case, but now using the 
 solution \eqref{ntimedomain} in the paraxial approximation in time-domain, we can 
construct  the general wavepacket for this solution
as
\begin{eqnarray}
 {\tilde n}_G(t,{\bf x})&=&{\tilde n}_0 \rho_z(t,z) \exp\left(i l \phi+i\frac{\omega^2+\omega_p^2}{2\omega}t\right)\times\nonumber\\
&&\int_0^\infty d\alpha\,  g(\alpha)  J_l\left(\alpha r\right)\exp\left(i\frac{S^2\alpha^2}{2\omega}t\right)\, . 
\label{wavepackettimedomain}
\end{eqnarray}
Of course, the general relativistic electron plasma wavepacket \eqref{wavepackettimedomain} is a solution of Eq.~\eqref{densityequationparaxial1}. It
presents acceleration in a longitudinal direction, while it remains arbitrary structured in the transversal directions. This structured form can be chosen at will by using different spectral functions. 
As examples of this wavepacket, let us calculate, in an analytical fachion, some simpler forms for different spectral factors.

{\it Case $g(\alpha)=1$, and $l=1$.} For this, we obtain that \cite{ryzhik}
\begin{equation}
\int_0^\infty d\alpha\,   J_1\left(\alpha r\right)\exp\left(i\frac{S^2\alpha^2}{2\omega}t\right)=\frac{1}{r}\left[1-\exp\left(i\frac{\omega r^2}{2 S^2 t}\right) \right]\, ,
\end{equation}
which produce a solution transversally decays as $1/r$, assuring a localization of it as it propagates in the $z$-direction. 
In Fig~\ref{figura2}(a), we plot the magnitude $|{\tilde n}_G(t,{\bf x})|^2$ for this solution, for the specific time $t=0.1$. Also $\omega/S^2=1/2$ (which plays the role of a time weighting), $u=0$, $v=0.1$, and $a=0.1$. We can see the transversal localization of this solution.

{\it Case $g(\alpha)=\alpha J_1(\lambda\alpha^2)$, and $l=1/2$.} In this, when $\lambda> S^2 t/(2\omega)$, then \cite{ryzhik}
\begin{eqnarray}
&&\int_0^\infty d\alpha\,  \alpha J_1(\lambda^2\alpha) J_{1/2}\left(\alpha r\right)\exp\left(i\frac{S^2\alpha^2}{2\omega}t\right)=\nonumber\\
&&\qquad\quad\sqrt{\frac{2}{\pi\lambda r^2}}\sin\left(\frac{\lambda r^2}{4 (\lambda^2-S^4 t/4\omega^2)} \right)\nonumber\\
&&\qquad\qquad\times \exp\left(i\frac{S^2 t r^2}{8 \omega (\lambda^2-S^4 t/4\omega^2)}\right)\, .
\end{eqnarray}
again, there is a $1/r$ transversal decay in the solution. In Fig.~\ref{figura2}(b), we plot  $|{\tilde n}_G(t,{\bf x})|^2$
for this solution, for the same values 
$t=0.1$, $\omega/S^2=1/2$, $u=0$, $v=0.1$,  and $a=0.1$. We also choose $\lambda=1$ to fulfill the condition to obtain this solution.

As it is clear, any other spectral function $g(\alpha)$
can be considered to construct this kind of localized accelerating solution.

\subsection{Generalized curved trajectory plasma wavepacket} 
\label{gencurvrajeplaswavep}

For this case, using solution \eqref{ntimedomainspace},
 it is also straightforward to show that the wavepacket, defined as
\begin{align}
    {\tilde n}_G(t,{\bf x})&= {\tilde n}_0  \exp\left(i k z+i\omega t\right) \rho_x(x,z) \rho_y(y,z)\nonumber\\
&\phantom{=}\qquad \times \int_0^\infty d\beta\,  g(\beta)\exp\left(- i \beta^2 z/2k\right)\, ,
\label{generalsolutionspacedomainwavepacket}
\end{align}
satisfies all the conditions of the paraxial approximation in space-domain. Anew, $g(\beta)$ is the arbitrary spectral function \citep{iwo} that allows the construction of different relativistic electron plasma wavepackets. This function will help to arbitrarily modulate the fall-off of the Airy function along the longitudinal direction $z$, as we show in the examples.

In this form, solution \eqref{generalsolutionspacedomainwavepacket} represents the most general form of propagation that shows parabolic trajectories in space. This Airy-like wavepacket propagating solution is a relativistic generalization of the one presented in Ref.~\cite{hehe}. 

Electron density wavepacket \eqref{generalsolutionspacedomainwavepacket} allows to have a whole broad family
of different structured wavepackets, as we demand that $\beta\neq 0$. On the contrary, by restricting ourselves to the $\beta=0$ choice, we recover the solution found in \cite{hehe}, but we rule out the possibility to define general wavepackets.

As examples, let us consider two simple wavepackets for different spectral functions.

\begin{figure*}
\subfloat[]{\includegraphics[width = 3.5in]{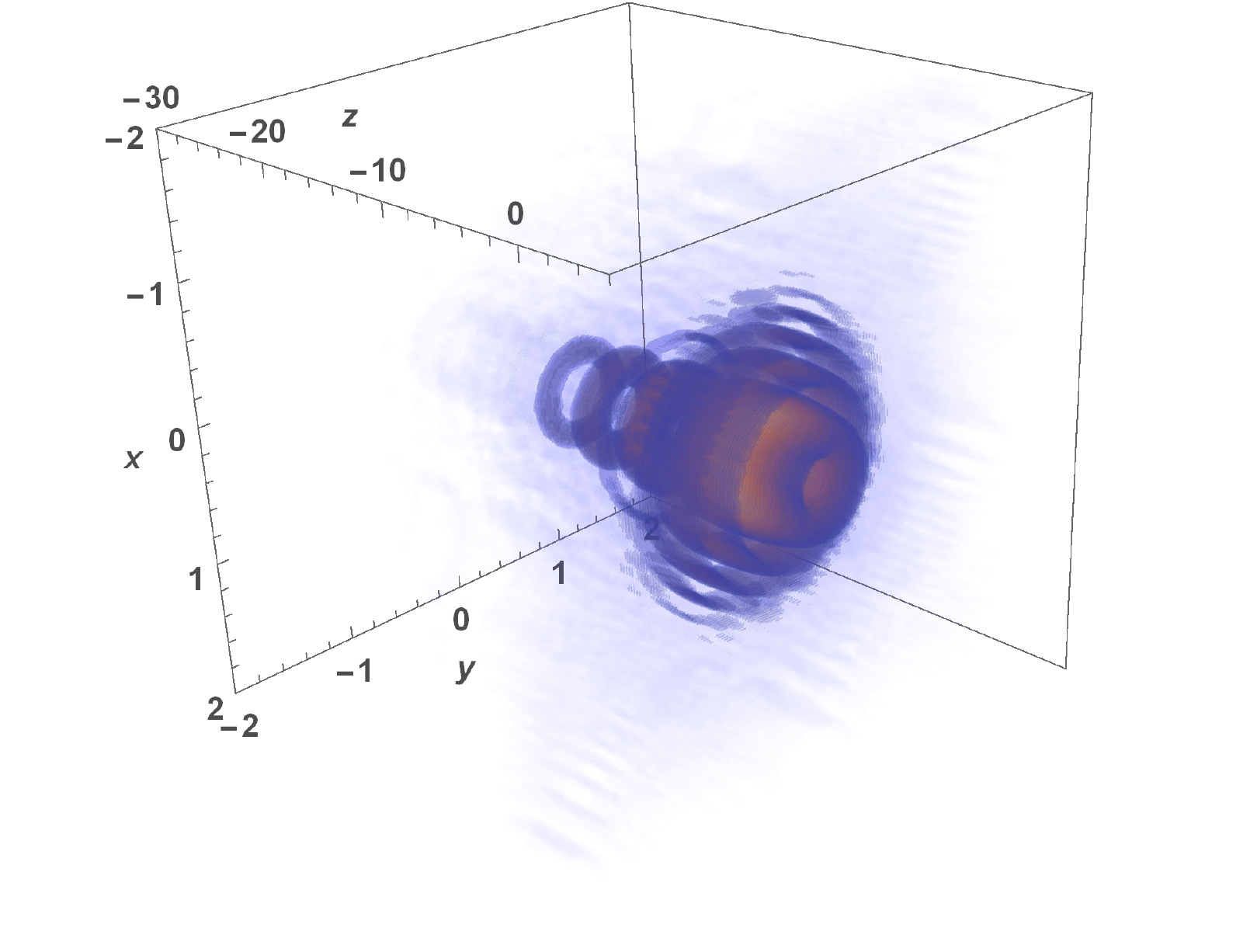}} 
\subfloat[] {\includegraphics[width = 3.5in]{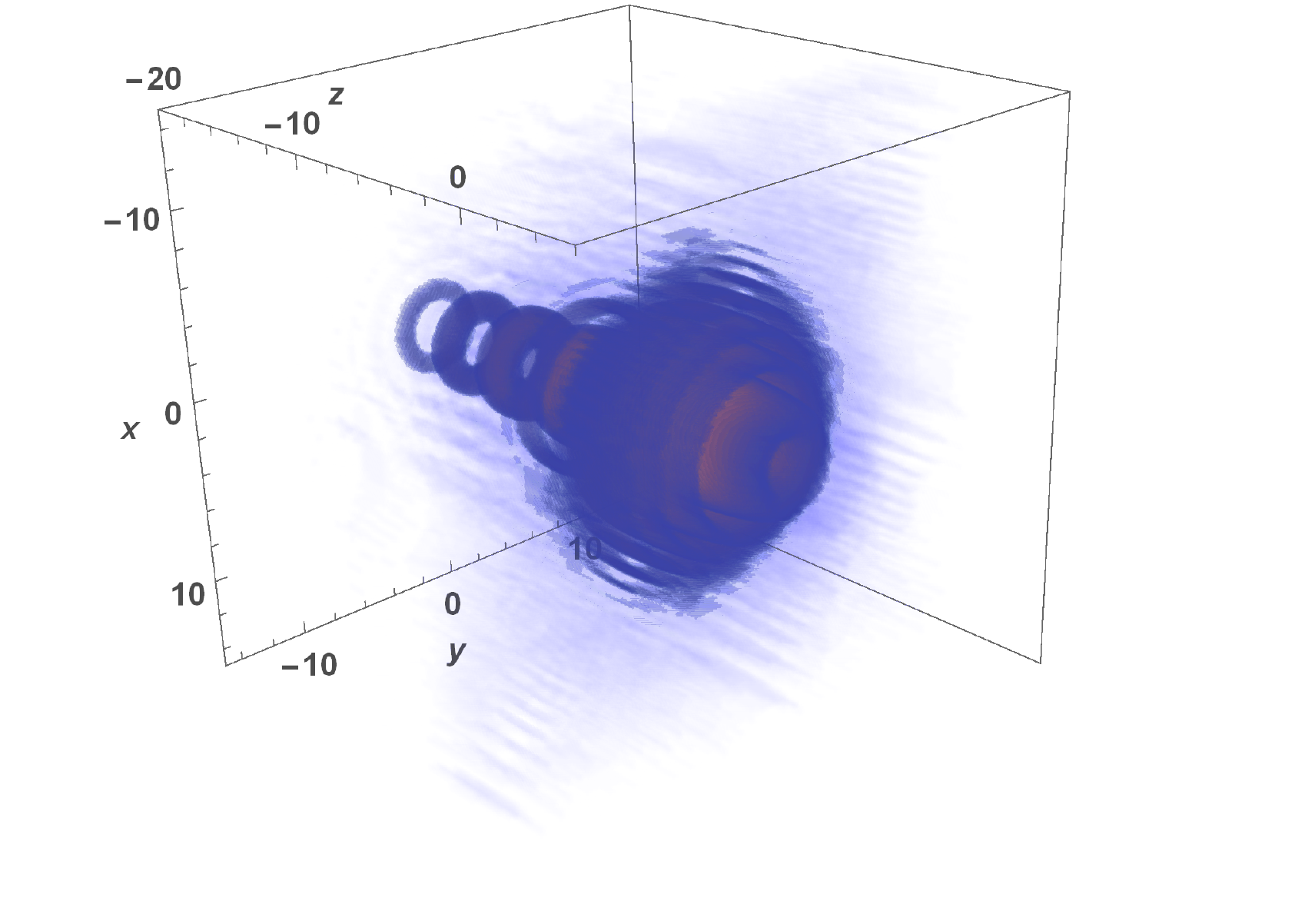}}\\
\subfloat[] {\includegraphics[width = 3.5in]{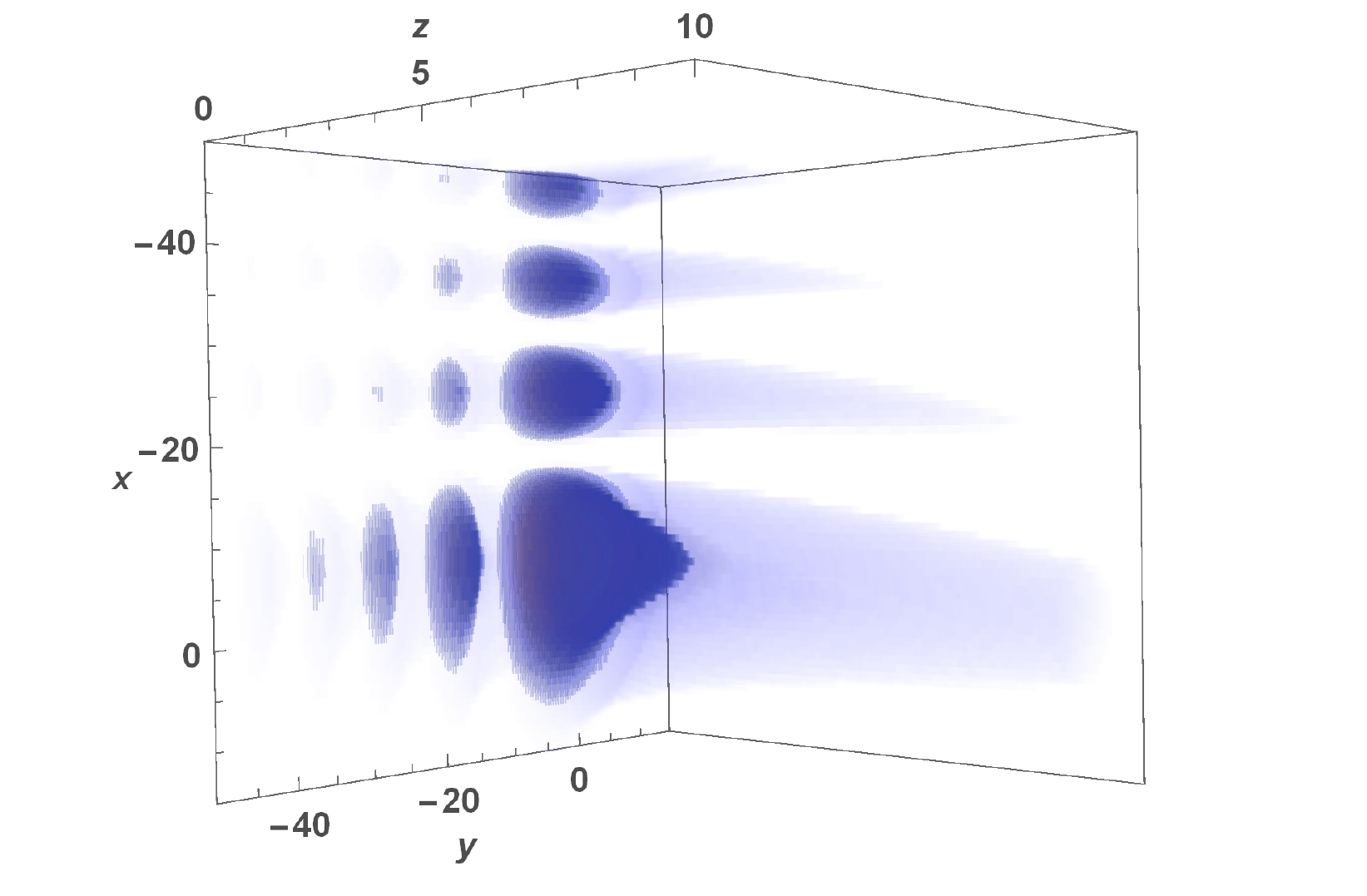}} 
\subfloat[] {\includegraphics[width = 3.5in]{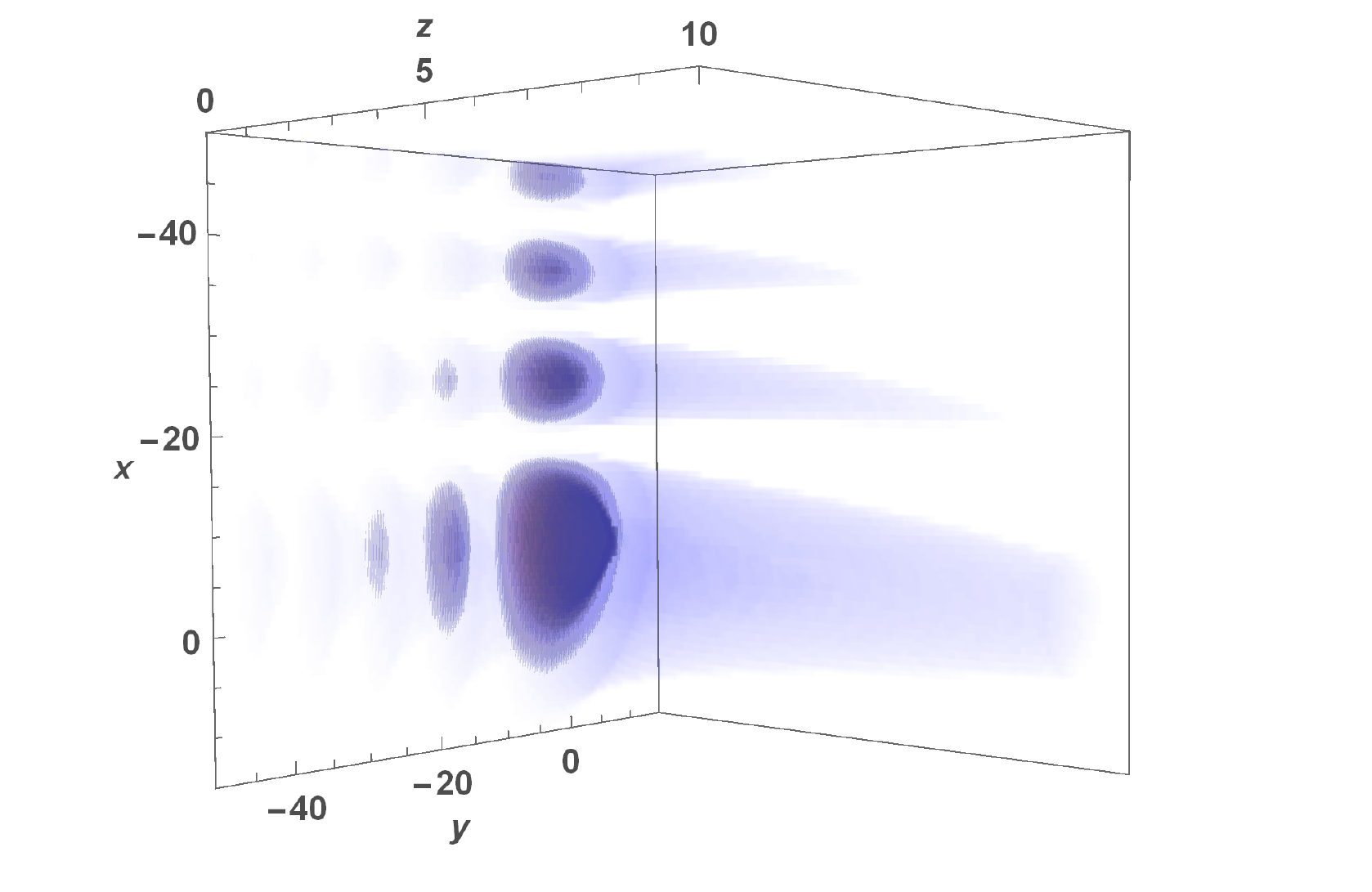}} 
\caption{Density plots for  $|{\tilde n}_G|^2$ of the generalized wavepacket solutions for different spectral funtions. Brighter colors stand for higher density values.  (a) Density \eqref{wavepackettimedomain} with spectral function $g=1$. (b) Density \eqref{wavepackettimedomain} with spectral function $g=\alpha J_1(\lambda \alpha^2)$. (c) Density \eqref{generalsolutionspacedomainwavepacket} with spectral function $g=1$. (d) Density \eqref{generalsolutionspacedomainwavepacket} with spectral function $g=\cos(1/\beta^2)$.}
\label{figura2}
\end{figure*}

{\it Case $g(\beta)=1$}. In this case, for $z>0$, we obtain \cite{ryzhik}
\begin{align}
\int_0^\infty d\beta\,  \exp\left(- i \beta^2 z/2k\right)=\sqrt{\frac{\pi k}{4 z}}(1+i)\, .
\end{align}
This wavepacket decays as $1/z$ along its propagation. In  Fig.~\ref{figura2}(c), we plot  $|{\tilde n}_G(t,{\bf x})|^2$ for this case, using 
$k=0.1$,  $u_x=0=u_y$, $v_x=0.1$, $v_y=0.3$, $a_x=0.1$ and $a_y=0.2$. We see the curved trajectories in the $x-z$ and $y-z$ planes
that different lobes follow, and its decays along $z$ direction.

{\it Case $g(\beta)=\cos(a^2/\beta^2)$}. In this case, for  $z>0$ and arbitrary $a>0$, we get \cite{ryzhik}
\begin{eqnarray}
&&\int_0^\infty d\beta\,\cos(a^2/\beta^2)  \exp\left(- i \beta^2 z/2k\right)=\nonumber\\
&&\sqrt{\frac{\pi k}{16 z}}(1-i)\left[\exp\left(-\sqrt{\frac{2a^2 z}{k}} \right)+\exp\left(-i\sqrt{\frac{2a^2 z}{k}} \right)\right]\, ,\nonumber\\
&&
\end{eqnarray}
which presents a faster decays along the $z$ direction of propagation. This can be seen in  Fig.~\ref{figura2}(d), where $|{\tilde n}_G(t,{\bf x})|^2$ for this solution is plotted, using   the same values than previous case, and $a=1$.  Again, curved trajectories are obtained.
Other curved trajectories waepackets can be constructed using different spectral functions.

Other general wavepacket solutions can be constrcuted from here. For example, we can integrate solution \eqref{ntimedomainspace} in $\omega$, to consider a wavepacket formed for all possible frequencies.

\section{Conclusions}

The relativistic electron plasma wave equation, followed by plasma density perturbations, allows to have solutions in terms of Airy
functions, in equivalence with other relativistic wave equations. These modes,  
very different from the usual plane wave solution, correspond to Airy-like wavepackets, that represent structured and non-spreading propagation of these relativistic electron plasma waves. 

We have shown Airy-like forms of propagation in an exact form \eqref{exactnairprogrpab} and in paraxial approximated forms \eqref{ntimedomain} and \eqref{ntimedomainspace}. These are the most general solutions of this kind for the perturbed density with dynamics given by Eq.~\eqref{densityequation}.  
The exact solution \eqref{exactnairprogrpab} shows that this mode has different high-intensity lobes and independent accelerations in the $y$ and $z$ directions, with respect to $\eta$ and $\xi$ directions. On the other hand, the paraxial propagation of Airy wavepackets can be studied from two different approximations: time-domain \eqref{ntimedomain} and space-domain \eqref{ntimedomainspace}. In the first case, the solution accelerates in time as it moves in space. This acceleration is present in a longitudinal direction, while it remains arbitrarily structured in the transversal direction. In the latter, wavepackets follow curved parabolic trajectories in space, that accelerate in the transverse direction.
These found solutions works as a base for more general wavepackets. We have shown how
these Airy-like solutions can be used to construct general classes of arbitrary wavepackets that will be dependent on different and arbitrary spectral functions. These solutions can be used to produce new kinds of three-dimensional structured  of electron plasma propagation. 
Therefore, in this paper, we have
presented the most general form for an Airy wavepacket in relativistic electron plasmas.

It is interesting to discuss the diverse applications of the different  solutions developed in this work. Mainly, all the above four-dimensional Airy solutions allow to manipulate the electron plasma density and energy along the non-constant path of propagation of the wave. Therefore, a manipulable localization of energy of the wave (accelerating in spacetime, in space, or in curved trajectories) can be achieved by properly choosing  the different parameters of the solutions, as the Poynting vector of the wave is no longer constant.   This can be applied, for example, to laser wakefield acceleration or 
to plasma propulsion, where by using these solutions, electrons can be accelerated at arbitrary magnitude and directions. This will be study in future works.

Finally, the same kind of procedure developed in this work, can be used to study similar Airy-like propagation for other plasma waves, such as electromagnetic plasma waves. In those cases, it is also expected that general wavepacket solutions can be constructed.

{\bf Author Contribution Statement:} All authors contributed equally to the paper.

{\bf Data Availability Statement:} No datasets were generated or analysed during the current study.

\begin{acknowledgements}
FAA thanks to FONDECYT grant No. 1230094 that supported this work.
 \end{acknowledgements}

\end{document}